# Exact Cancellation of Emittance Due to Coupled Transverse Dynamics in Solenoids and RF Couplers

David H. Dowell, Feng Zhou, and John Schmerge
SLAC National Accelerator Laboratory

**Abstract**

Small, stray magnetic and RF fields in electron guns and injectors can perturb an electron beam and introduce correlations between the otherwise orthogonal transverse trajectories.  These correlations couple the x and y dynamics which increases the transverse emittance.  If the correlation becomes 'diluted' or randomized in the beam transport then as the correlation disappears the 4D emittance increases.  This paper discusses two important correlations common to most electron injectors. The first results from the coupling of a weak quadrupole field with beam rotation in a solenoid, and the second x-y coupling is generated by an asymmetric on-axis RF field due to a high-power RF coupler or cavity port.  This paper shows that a small quadrupole field combined with solenoidal focusing can result in significant emittance growth due to coupled transverse dynamics.  It also shows how adding a skewed quadrupole field can exactly cancel this correlation and its emittance. Similar emittance cancellation is demonstrated for asymmetric RF fields, with the degree of cancellation limited by the electron bunch length.  Analytic expressions are derived and compared with simulations and experiments.

## I. INTRODUCTION

Solenoids are ubiquitous in electron guns used for microscopy, injectors for high energy accelerators, and other applications requiring high-quality electron beams.  Solenoids are used to focus and control the beam's size as it's accelerated from the cathode and to focus the beam onto a sample or into a high-energy accelerator.  Although usually considered benign, focusing with a solenoid can generate three major aberrations: chromatic, geometric, and coupled transverse dynamics aberrations [1].  The chromatic and geometric aberrations are well-known and their effects are included in most particle tracking and simulation codes.  On the other hand, coupled transverse dynamics is not as well-known and is not usually included in most simulations. The coupled-transverse dynamics of the beam in a solenoid in conjunction with a weak, stray quadrupole field is the first topic of this paper.

The second topic is the aberration due to asymmetric transverse fields of RF couplers used in both normal and superconducting accelerators.  This work shows the coupler's fields produce an instantaneous transverse voltage kick like that of a rotated quadrupole.  It is also demonstrated that weak skewed DC quadrupole fields can be used to reduce most of the RF coupler emittance.  The technique is effective provided the bunch sees the coupler fields as frozen in time during its transit and the bunch is relatively short.  Calculations and simulations verify the bunch length contribution to the coupler emittance is negligible for the LCLS-II injector parameters.  The emittance is mostly due to skewed transverse fields.

## II. THE SOLENOID'S QUADRUPOLE FIELD

There are quadrupole fields at the entrance and exit of all solenoids and these fields are related to the solenoid's fringe field.  This is most easily seen by writing Maxwell's equation for the divergence of the magnetic field, $\vec{\nabla} \cdot \vec{B} = 0$, in cylindrical coordinates

$$\frac{1}{r}\frac{\partial}{\partial r}(rB_r) + \frac{1}{r}\frac{\partial B_\theta}{\partial \theta} = -\frac{\partial B_z}{\partial z} \tag{1}$$

Thus, the sum of the radial and azimuthal field gradients equals the slope of the longitudinal field. And therefore, the radial and azimuthal fields are at the ends of the solenoid because that's where the greatest slope is. Multiplying by $dz$ and integrating from $-\infty$ to the magnet center at z=0 gives

$$\int_{-\infty}^{0}\left(\frac{1}{r}\frac{\partial}{\partial r}(rB_r)+\frac{1}{r}\frac{\partial B_\theta}{\partial \theta}\right)dz = -\int_{-\infty}^{0}\frac{\partial B_z}{\partial z}dz = -\int_{0}^{B_0}dB_z = -B_0 \quad (2)$$

Here $B_0$ is the peak field at the center of the magnet. In the hard-edge model the peak magnetic field is assumed constant over an effective length such that their product is equal the actual field integral. Defining this length to be $L_{fringe}$ and assuming the transverse fields do not depend upon z over this effective length, one can write the equation for the transverse field gradients at the solenoid entrance as,

$$\left(\frac{1}{r}\frac{\partial}{\partial r}(rB_r)+\frac{1}{r}\frac{\partial B_\theta}{\partial \theta}\right)_{entrance} = -\frac{B_0}{L_{fringe}} \quad (3)$$

The equation for the exit field gradients is the same except with the opposite sign,

$$\left(\frac{1}{r}\frac{\partial}{\partial r}(rB_r)+\frac{1}{r}\frac{\partial B_\theta}{\partial \theta}\right)_{exit} = \frac{B_0}{L_{fringe}} \quad (4)$$

Hence, the sum of the radial and azimuthal integrated field gradients at each end should be equal in magnitude but have opposite signs. Therefore, one would naively think the fringe field kicks given to the beam should cancel. However, because the beam rotates in the solenoid, it 'sees' the exit quadrupole field with a different skew than the opposite of the entrance quadrupole fringe field skew. Thus, in general the fringe field kicks don't cancel and the transverse emittance can increase.

Figure 1 shows magnetic measurements for the LCLS-I gun solenoid [2]. These rotating coil measurements illustrate the 90-degree phase change or polarity reversal between entrance and exit fringe fields as given by Eqns. (3) and (4). In addition, the measurements also show small but, as will be seen, surprisingly significant quadrupole strength in the fringe fields. These measurements show the quadrupole fields peak at the ends of the solenoid. This is because all the transverse fields, like the quadrupole field, scale with $\frac{dB_z}{dz}$ which is its largest with opposite polarities at the ends.

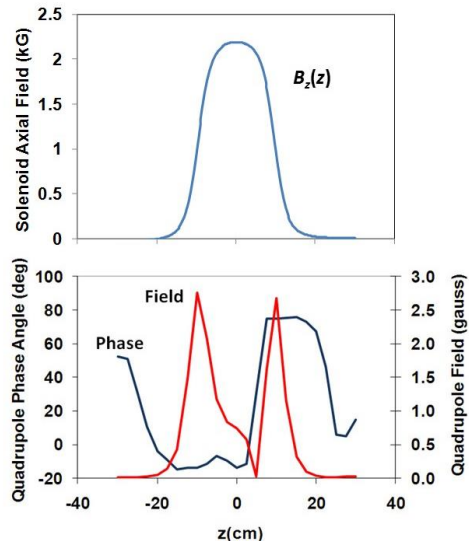

FIG. 1(color): Magnetic measurements of the LCLS gun solenoid with the integrated solenoid field at 0.046 T-m. Top: Hall probe measurements of the solenoid axial field. The transverse location of the measurement axis (the z-axis) was determined by minimizing the dipole field. Bottom: Rotating coil measurements of the quadrupole field. The rotating coil dimensions were 2.5 cm long with a 2.8 cm radius. The measured quadrupole field is thus integrated over these dimensions.

It is assumed that small asymmetries in the solenoid's coil windings or permeability variations of the magnet yoke and its environment can perturb the large radial and longitudinal fields and result in the small quadrupole fields shown in Figure 1. However, in our experience, the sources of these small fields were difficult to verify and control even with state-of-the-art finite-element-analysis calculations. Therefore, we decided to install weak normal and skew quadrupole correctors in the LCLS-I solenoid and optimize their settings with the beam itself.

The azimuthal fields can be expressed as a multipole expansion in the azimuthal coordinate,

$$B_\theta = B_1 \cos\theta + B_2 \cos 2\theta + B_3 \cos 3\theta + \cdots \qquad (5)$$

Here $B_1$ is the dipole component of the azimuthal field, $B_2$ is the quadrupole component and $B_3$ is the sextupole. Proper mechanical alignment and operational techniques, such as beam-based alignment, are effective at eliminating the dipole term. Therefore, the dipole field will be ignored in this analysis. However, such techniques are ineffective at mitigating the quadrupole and higher-order multipoles. In addition, few of the beam simulations used to optimize the emittance include these fields in their calculations. The next section of this paper presents an analysis of the solenoid's quadrupole fringe fields by first describing the beam's dynamics in the skewed quadrupole and solenoid fields, and second by showing these dynamics can be exactly canceled with a weak, skewed quadrupole field.

### III. THE COUPLING OF TRANSVERSE DYNAMICS IN A SOLENOID

Both the skewed quadrupole fields and the solenoid axial field couple the x-x' and the y-y', or r- and $\theta$-coordinates, and therefore can increase the emittance. What is important to note is that the effect begins with a quadrupole field weakly focusing the beam before it enters the solenoid. Even a quadrupole with no skew and by itself would generate no emittance, becomes 'skewed' by the rotation in the solenoid and a source of transverse emittance.

As described in the last section, the transverse coupling imposed on the beam by the entrance fringe field is exactly canceled by the exit fringe field. However, this cancellation is ruined by the rotation of the beam in the solenoid which introduces an additional skew angle to the beam. Without the rotation in the solenoid the two quadrupole fringe field kicks would cancel and there would be no emittance due to the quadrupole fringe fields of the solenoid.

Theoretical investigations describe the coupling dynamics by generalizing the Courant-Snyder theory with a 4D symplectic rotation [3]. Ripken [4] was the first to use this approach as summarized by Wiedemann [5]. In this analysis, the full 4D transverse transport matrix conserves the 4D emittance since the transformation is linear in 4-dimensions. However, the 2D sub-spaces of xx' and yy' gain emittance when the 4-D distributions are projected down to 2D. The linear 4D-transformation generates correlations between off-diagonal beam matrix elements such as xy, x'y, x'y', etc. which are normally zero. A skewed quadrupole can produce these cross terms and can be used to control these terms. Indeed, a later section on RF coupler fields shows that Maxwell's equations establish two relations between the four elements of the voltage kick matrix, and that with these symmetries the voltage kick matrix can be written as a beam rotation, like that produced by a skewed quadrupole field.

This section describes a mathematical model for the quadrupole-solenoid-quadrupole system and derives an expression for the emittance.

#### A. Analytic model for the quadrupole effect in a solenoid

Fig. 2 shows the mathematical model used to analyze the solenoid and quadrupole system. The entrance quadrupolar fringe field is modeled as a quadrupole with a skew angle $\alpha_1$ and focal length $f_1$, or an integrated quadrupole field of $L_{fringe} \frac{\partial B_y}{\partial x}\big|_{x,y=0}$ located at the entrance of a solenoid which rotates the beam the angle KL. The solenoid is followed by another quadrupole with skew angle $\alpha_2$ and focal length $f_2$.

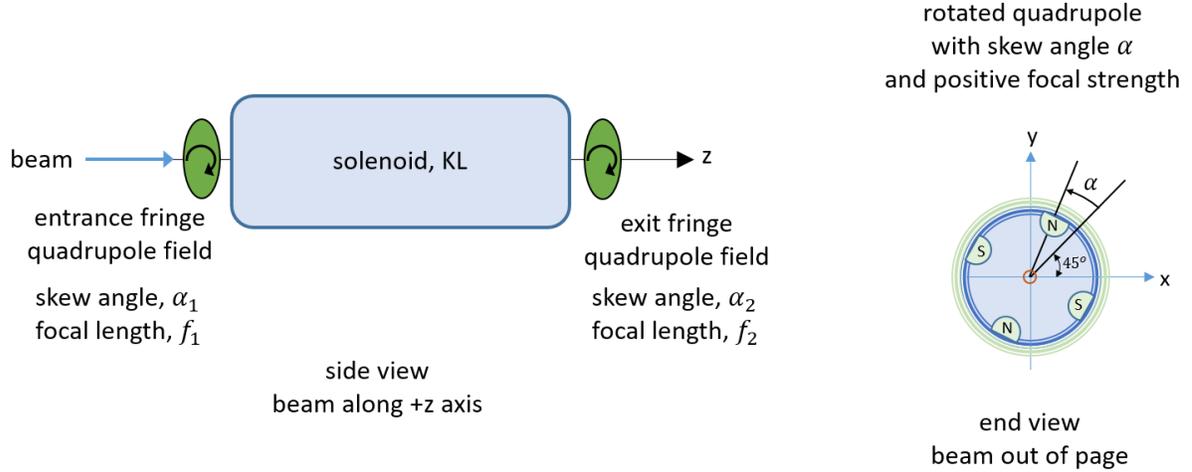

FIG. 2(color). Model of a solenoid with the quadrupole fields at the ends. Side view: The electrons are traveling in the +z direction (side view). End-view: The quadrupole polarity is positive and focusing for electrons if the N-pole is in the first-quadrant of the xy-plane. The quadrupole is a 'normal quadrupole' if the N-pole is at 45 degrees with respect to the x-axis or skew angle $\alpha = 0$. A 'skew quadrupole' has $\alpha = 45$ degrees which places the N-poles on the y-axis and the S-poles along the x-axis.

As described earlier, the fringe fields the solenoid's ends are equal in strength but have opposite polarities. The polarity reversal implies a 90-degree difference in quadrupole phase angle between the two fringe fields,

$$\alpha_2 = \alpha_1 + \frac{\pi}{2} \quad (6)$$

This phase shift of 90-degrees occurs near the solenoid center as verified by the quadrupole phase plot in Fig 1. It was also shown that the magnitudes of the fringe field focal lengths are equal,

$$f_2 = f_1 \quad (7)$$

These relations assume the focal length is always positive and the polarity of the field is determined by the skew angle. Thus, just two parameters: a skew angle, and a focal length (or the integrated field gradient), fully determine the quadrupole field at both ends of the solenoid.

### B. Emittance due to the quadrupolar fringe field of a solenoid

The emittance due to a quadrupole focusing through the rotation in the solenoid is computed assuming the entrance quadrupole field skew angle is zero and the field of the exit quadrupole is zero. This result is easily generalized to include the exit quadrupole and non-zero skew angles. It is important to note that in this model the 4D emittance remains zero however the emittances for the 2D phase spaces spanned by xx' and yy' increase. This is relevant for applications such as XFELs and UED where it is essential that the sub-space emittances be preserved and any correlations between the sub-spaces be removed.

The linear (x,x',y,y') phase space transformation of the beam through a thin quadrupole lens followed by a solenoid can be written as [6],

$$R_{sol}R_{quad} = \begin{pmatrix} \cos^2 KL & \frac{\sin KL}{K} & \sin KL \cos KL & \frac{\sin^2 KL}{K} \\ -K \sin KL \cos KL & \cos^2 KL & -K\sin^2 KL & \sin KL \cos KL \\ -\sin KL \cos KL & -\frac{\sin^2 KL}{K} & \cos^2 KL & \frac{\sin KL \cos KL}{K} \\ K \sin^2 KL & -\sin KL \cos KL & -K \sin KL \cos KL & \cos^2 KL \end{pmatrix} \begin{pmatrix} 1 & 0 & 0 & 0 \\ -\frac{1}{f_1} & 1 & 0 & 0 \\ 0 & 0 & 1 & 0 \\ 0 & 0 & +\frac{1}{f_1} & 1 \end{pmatrix} \quad (8)$$

Here $L$ is the effective length of the solenoid, $K \equiv \frac{eB_0}{2p}$, $B_0$ is the interior peak axial magnetic field of the solenoid, and $f_1$ is the focal length of the quadrupole field located at the entrance to the solenoid. The quadrupole field's skew angle is zero and the beam is rotated through the angle $KL$ by the solenoid.

Using Eqn. (8), the initial 4×4 beam matrix, $\Sigma(0)$, is transported through the quadrupole and solenoid producing the exit beam matrix, $\sigma(1)$,

$$\Sigma(1) = R_{sol}R_{quad}\, \Sigma(0) \left(R_{sol}R_{quad}\right)^T \quad (9)$$

The initial beam is assumed to be collimated with perfectly parallel rays and zero emittance. In this case, the initial beam matrix is

$$\Sigma(0) = \begin{pmatrix} \Sigma_{xx}(0) & 0 & 0 & 0 \\ 0 & 0 & 0 & 0 \\ 0 & 0 & \Sigma_{yy}(0) & 0 \\ 0 & 0 & 0 & 0 \end{pmatrix} \quad (10)$$

Where the two non-zero beam matrix elements are the beam sizes at the solenoid entrance squared,

$$\Sigma_{xx}(0) = \sigma_{x,sol}^2 \quad \text{and} \quad \Sigma_{yy}(0) = \sigma_{y,sol}^2 \quad (11)$$

The 4D-emittance is given by the 4x4 beam matrix

$$\epsilon_{n,4D} = \beta\gamma\sqrt{\det \Sigma(1)} \quad (12)$$

which is equal to zero, since the full beam matrix is symplectic in 4-dimensions and the initial 4D-emittance was zero. However, the xx' phase sub-space represented by a 2D sub-matrix is not conserved and its emittance increases. The xx' phase space emittance is given by the determinate of the sub-matrix,

$$\epsilon_{n,x} = \beta\gamma\sqrt{\det \begin{vmatrix} \Sigma_{xx} & \Sigma_{xx'} \\ \Sigma_{xx'} & \Sigma_{x'x'} \end{vmatrix}} \quad (12)$$

Beginning with $\sigma(1)$ in Eqn. (8), and working through tedious matrix algebra and applying some trigonometric identities gives the surprisingly compact result for the xx'-emittance for a normal quadrupole followed by a solenoid,

$$\epsilon_{n,quad+sol} = \beta\gamma\frac{\sigma_{x,sol}\sigma_{y,sol}}{f_1}|\sin 2KL| \quad (13)$$

Figure 3 compares this simple formula with a particle tracking simulation for a solenoid with a weak quadrupole field. The initial beam had zero emittance, zero energy spread and was circular and uniform. No space charge forces are included in the simulation. The figure shows the normalized emittances given by Eqn. (13) and the simulation plotted as a function of the rms beam size at the solenoid entrance. The quadrupole focal length is 50 meters at 6 MeV which is approximately the same as measured for the LCLS solenoid. Both the analytic theory and the simulation assume a short quadrupole field only at the solenoid's entrance. The simulation is slightly larger since it includes both this quadrupole effect and the geometric aberration described above. The good agreement verifies the model's basic assumptions and illustrates how even a very weak quadrupole field can strongly affect the emittance when combined with the rotation in a solenoid field.

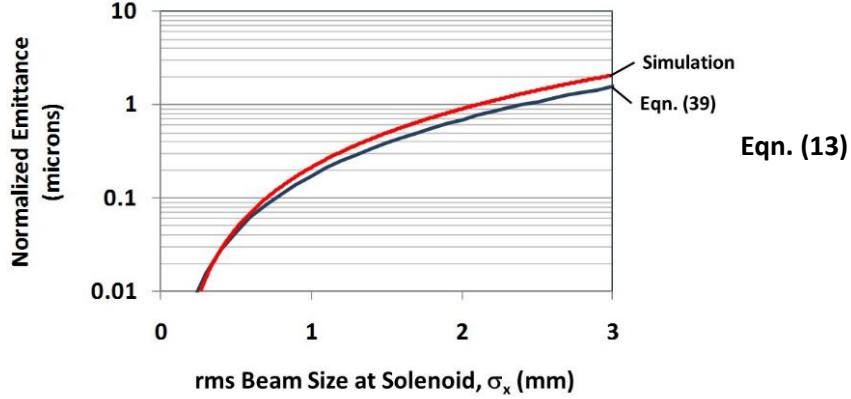

FIG. 3(color). Comparison of the emittance due to the quadrupole-solenoid coupling given by Eqn. (13) with a particle tracking simulation for the case of the LCLS solenoid. For a beam energy of 6 MeV the quadrupole focal length was 50 meters and the solenoid had an integrated field of 0.046 T-m. The simulation (solid red) is done with the GPT code [7].

This normalized emittance is seen to be independent of beam energy by expressing it in terms of the integrated quadrupole field instead of the quadrupole focal length,

$$\epsilon_{n,quad+sol} = \frac{e\sigma_{x,sol}\sigma_{y,sol}}{mc} L_{fringe} \left.\frac{\partial B_y}{\partial x}\right|_{x,y=0} |\sin 2KL| \quad (14)$$

This expression is for a quadrupole plus solenoid if the entrance quadrupole field isn't skewed. If the field is skewed an angle $\alpha_1$ as shown in Figures 1 and 4 then the emittance is given by adding the quadrupole's skew angle to the solenoid's rotation angle,

$$\epsilon_{n,quad+sol}(\alpha_1) = \frac{e\sigma_{x,sol}\sigma_{y,sol}}{mc} L_{fringe} \left.\frac{\partial B_y}{\partial x}\right|_{x,y=0} |\sin 2(KL + \alpha_1)| \quad (15)$$

Figure 4 compares Eqn. (15) with a simulation for a 50-meter focal length quadrupole followed by a strong solenoid (focal length of ~15 cm). The emittance is plotted as a function of the quadrupole angle of rotation. In both the analytic theory and the simulation, the emittance becomes zero whenever

$$KL + \alpha_1 = \frac{\pi}{2} \quad (16)$$

The slight offset in magnitude between the theory and simulation could result from the 3rd-order emittance of the solenoid's non-linear radial fringe fields which is included in the simulation but not in the model.

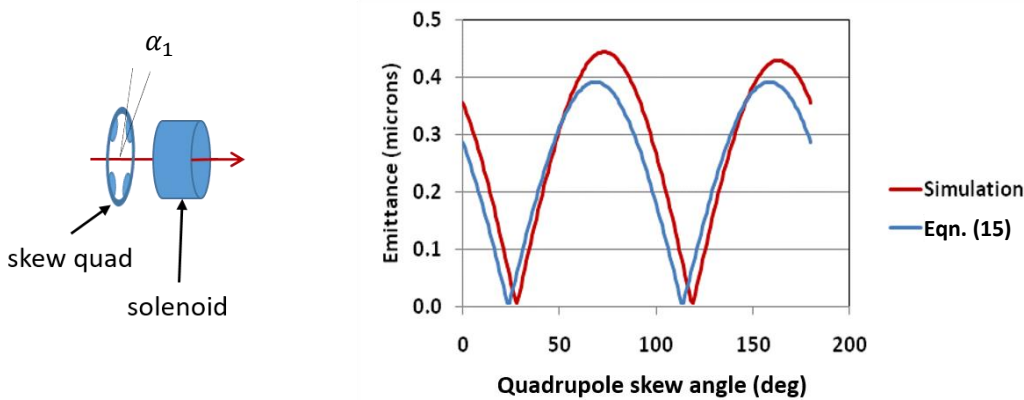

FIG. 4(color).  The emittance for a quadrupole-solenoid system (shown on the left) plotted as a function of the quadrupole rotation angle.  The theory emittance (solid blue) is computed using Eqn. (15) and the simulation (solid red) is done with the GPT code.  The beam size at the solenoid is 1 mm-rms for both the x- and y-planes.

The emittance for the complete solenoid with both the entrance and exit quadrupole fringe fields can now be written with, the aid of Eqns. (6) and (7), as

$$\epsilon_{n,quad+sol} = \frac{e\sigma_{x,sol}\sigma_{y,sol}}{mc} L_{fringe} \left.\frac{\partial B_y}{\partial x}\right|_{x,y=0} |\sin 2(KL + \alpha_1) - \sin 2\alpha_1| \quad (17)$$

Here $L$ is the effective length of the solenoid, $K \equiv \frac{eB_0}{2p}$, $B_0$ is the interior peak axial magnetic field of the solenoid, and the x- and y-rms beam sizes at the entrance to the solenoid are $\sigma_{x,sol}$ and $\sigma_{y,sol}$, respectively.  The beam is rotated through the Larmor angle $KL$ by the solenoid as described earlier.

It is relevant to point out some of the features of Eqn. (17).  First consider the situation when both quadrupoles are perfectly aligned without any skew, i.e. $\alpha_1 = 0$.  Then there is no emittance contribution from the exit quadrupole since it's a normal quad, however, even with zero skew angle the entrance quadrupole appears skewed by the beam due to its rotation in the solenoid.  And therefore, the emittance increases.  In this case, $\alpha_1 = 0$, the emittance is unaffected by the strength of the exit fringe field although it will give the beam a 'quadrupole-like' shape aligned to the x- and y-axes.   This is not true when $\alpha_1 \neq 0$.  Eqn. (17) then shows that if the entrance quadrupole field is skewed, both quadrupole fields and the solenoid rotation increase the emittance and the final emittance depends upon the polarity of the solenoid field.  This is distinctly different from the expected same emittance for both polarities due to the focal strength being proportional to the B-field squared.

In addition, Eqn. (17) suggests that adding the appropriately skewed quadrupole field near the solenoid cancels the effect and allows complete recovery of the initial emittance.  This is because the fields produce this emittance with linear correlation which can be canceled with the appropriately skewed quadrupole correction field.  Details of this cancellation are described next.

### C.  Cancellation of the emittance due to transverse coupling in a solenoid

This section shows that if the stray and distorting fields can be described as quadrupole multipoles rotated about the beam axis, then the transverse coupling emittance can be exactly cancelled with a skewed quadrupole.  The concept adds a correction quadrupole after the solenoid with the skew and focal length necessary to cancel the skew produced by the beam's rotation in the solenoid.  In this section, the solenoid and quadrupole fields are modeled in a quadrupole+solenoid+quadrupole (qsq) configuration, and the fringe field prescription abandoned for the more general and flexible configuration. The solenoid's fringe field configuration is recovered by relating the entrance and exit field skew angles as described earlier. The qsq configuration is used in comparing the analytic theory with the numerical simulation.

The previous discussion showed the transverse coupling emittance depends upon the square of the beam size. Therefore, the focal strength of the correction quadrupole depends inversely upon the beam size at its location. This means that half the beam size will require a 4-times stronger integrated quadrupole field strength for the corrector. And since the beam is converging inside the solenoid, the beam size at a correction quadrupole located near the solenoid exit will be smaller than the size at the solenoid's entrance. This additional effect can be included in the theory by defining the x and y rms beam sizes at the quadrupole as $\sigma_{x,quad}$ and $\sigma_{y,quad}$, respectively, and revising the expression for the transverse-coupled emittance. Then the emittance for the qsq configuration can be written as,

$$\epsilon_{n,qsq} = \frac{e}{mc} \left| \sigma_{x,sol} \sigma_{y,sol} Q_1 \sin 2(KL + \alpha_1) + \sigma_{x,quad} \sigma_{y,quad} Q_2 \sin 2\alpha_2 \right| \tag{18}$$

This expression is valid for short quadrupoles with integrated quadrupole field, $Q_1$, and skew angle, $\alpha_1$, at the entrance of the solenoid and a correction quadrupole with $Q_2$ and $\alpha_2$ at the exit of the solenoid as shown in Fig. 5. It is important to include the beam sizes since simulations show the beam size at the quadrupole is half that at the entrance of the solenoid which would reduce the quadrupole emittance by a factor of two-squared or four.

The Q's are integrals of the quadrupole fields and are defined in terms of an effective length and sextupole gradient for each end of the solenoid,

$$Q_{1,2} = L_{1,2} \frac{\partial B_y}{\partial x}\bigg|_{1,2} \tag{19}$$

where the subscript 1,2 refer to the entrance and exit quadrupoles, respectively. For example, $L_1$ is the effective length of the entrance quadrupole. The field gradient is evaluated at beam center, x,y=0. The integrated quadrupole field, Q, in terms of the quadrupole focal length, $f_q$, and beam momentum, $p_0$, is

$$Q = \frac{p_0}{e} \frac{1}{f_q} = \frac{\beta \gamma mc}{e} \frac{1}{f_q} \tag{20}$$

The layout and parameters given in Fig. 5 are used below to compare the analytic model with the numerical simulation of a commercial electron beam tracking code [7]. As discussed earlier, the beam size at $Q_2$ is approximately half its the size at $Q_1$. Hence, we have made the $Q_2$-quadrupole twice the length and its focal length less than half of $Q_1$ to makeup this factor of four and cancel the emittance produced by the first quadrupole and solenoid at similar quadrupole field gradients. The Q2 beam size for the analytic model is taken from simulation.

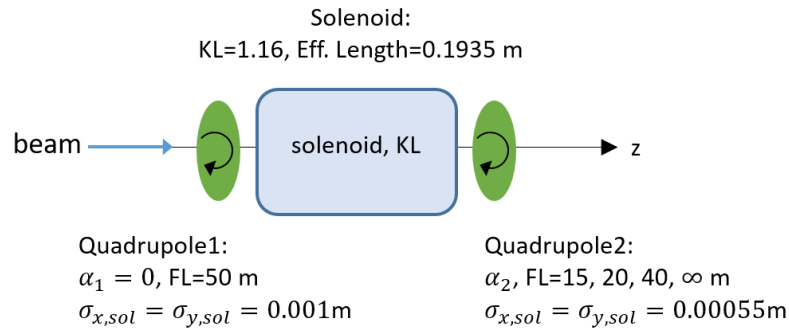

FIG. 5(color). Quadrupole-solenoid-quadrupole configuration and parameters used for comparing the analytic model and numerical simulation.

Figure 6 compares emittance calculations of the analytic theory and the numerical simulation for the qsq-configuration with parameters given in Fig 5. The plots show the normalized x-plane emittance for a 6 MeV beam as a function of the skew angle of the Q2-quadrupole for theory (left) and simulation (right). In all cases, the Q1-quadrupole focal length is 50 m and its skew angle is zero. The emittance is plotted for Q2 focal lengths of 15, 20, 40 and infinity. The emittance with infinite Q2 focal length is the transverse-coupled emittance generated by the uncorrected Q1-quadrupole and the solenoid combination or qs configuration.

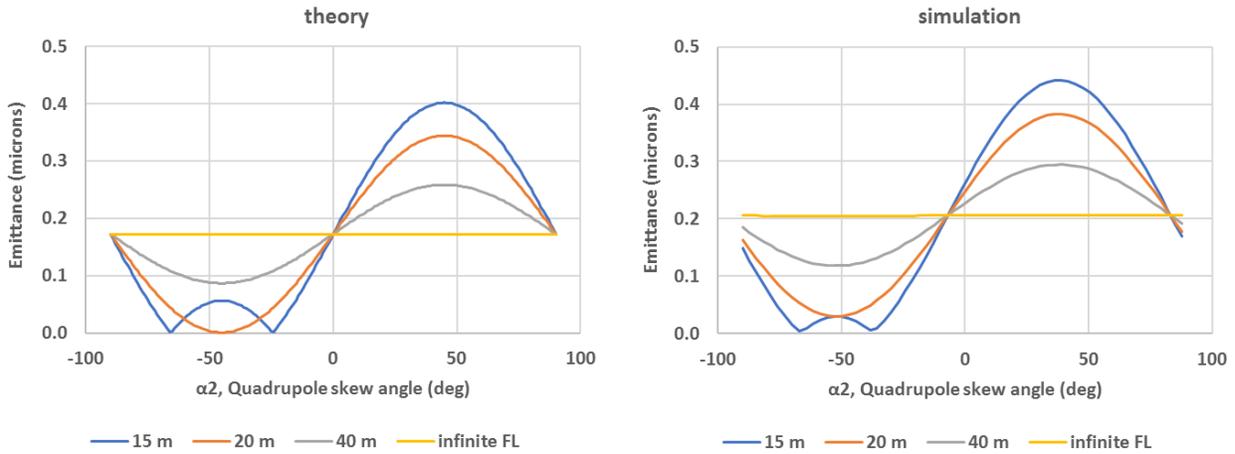

FIG. 6(color). Comparison of the traverse-coupled emittance for a qsq-configuration as a function of the Q2 skew angle as computed by theory (left) and a numerical simulation (right). The configuration and parameters used in the calculations are given in Fig. 5. The emittance is shown for Q2-quadrupole focal lengths of 15, 20, 40 and infinite meters.

Both the theory and simulation shown in Figure 5 indicate the Q2-quadrupole focal length needs to be shorter than 20 meters to cancel the emittance produced by the 50-meter focal length of Q1-quadrupole plus solenoid. As discussed earlier with Eqn. (18), this is mostly because the beam size is smaller at Q2 than at Q1.

### D. Simulation and measurements with quadrupole field in a solenoid

The effectiveness of the quadrupole correction has been validated with the simulations of the LCLS injector. The LCLS injector includes 5.5-MeV RF gun, a main solenoid for emittance compensation, and two linac sections to boost energy to 135 MeV [8]. The ASTRA code [9] including 3D space-charge forces is used to simulation the LCLS injector. Figure 7 shows the emittance vs. z for no quadrupole fields with only the solenoid field (blue), the emittance due to a weak, normal skew quadrupole at the entrance of the solenoid (green) and the weak quadrupole emittance corrected by a skewed quadrupole after the solenoid (red). These results show the emittance grows about 20% if there is a weak quadrupole field situated at the entrance of the solenoid. The red curve shows this emittance growth can be completely corrected by a weak skewed quadrupole in the region downstream of the solenoid. Thus, the simulations verify the analyses of the previous sections.

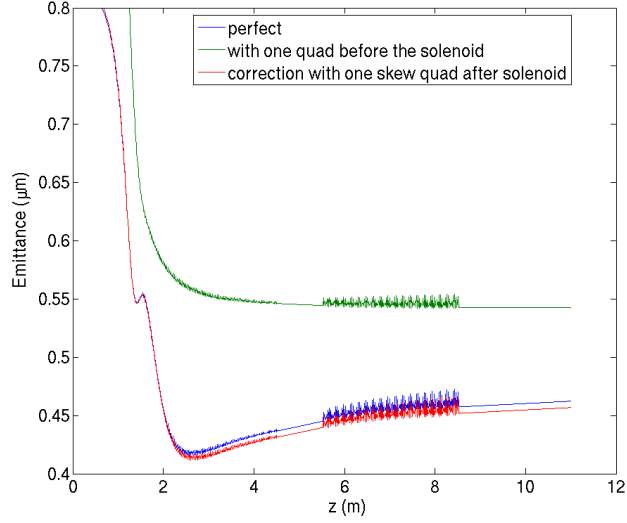

FIG. 7. Simulations of emittance correction with a skewed quadrupole correction: The emittance without a quadrupole field at solenoid entrance (blue), the emittance with a normal quadrupole at the entrance of the solenoid (green), and emittance correction with a skew quadrupole downstream of the solenoid (red).

Two long quadrupole correctors, one normal and one skewed, are installed inside of the solenoid for the LCLS-I injector. Figure 8 shows one example of the experimental data of the injector emittance measured at 135 MeV vs. the normal quadrupole corrector field strength. Shown are typical measurements made during LCLS-I operations to optimize the beam emittance. The measured emittance is plotted as a function of the normal quadrupole corrector field. Variation of a normal quadrupole field by itself should produce no emittance. However, because the normal field combines with other quadrupole fields which are skewed, it can affect the total skew angle and field amplitude of all the local quadrupole fields, and thereby change the emittance.

The data in Figure 8 have been fit with the quadratic sum of the quadrupole-solenoid-quadrupole emittance with the other miscellaneous emittances, $\epsilon_{other}$, using the following expression,

$$\epsilon_{expt} = \sqrt{\epsilon_{other}^2 + \left(\epsilon_{quad1-sol} + bQ_2\right)^2}$$

In this model, the contribution to the emittance by the transverse-coupled dynamics is all in the parameter $\epsilon_{quad-sol}$. And the emittance produced by the second quadrupole after the solenoid is $bQ_2$ where $Q_2$ is the integrated quadrupole field given by Eqn. (19), $b$ is a parameter related to the beam size and the skew angle determined from the fit,

$$b \equiv \frac{e}{mc} \sigma_{x,quad} \sigma_{y,quad} \sin 2\alpha_2$$

The fit shown by the green-dashed curve gives $\epsilon_{other}$ =0.45 microns and $\epsilon_{quad-sol}$ =0.11 microns. The fit has been made to both the x- and y-plane emittances for simplicity. Although the offset between the x- and y-planes emittances suggest the other-effects emittance for the y-plane is 25 nm larger than the x-plane's other-emittance. The data is shown with 5% error bars.

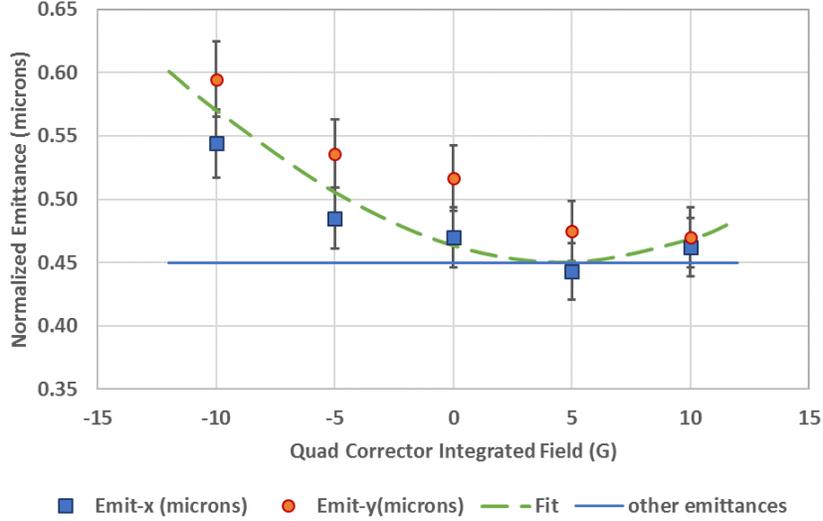

FIG. 8. Measured LCLS injector emittance vs. weak quadrupole. The beam energy is 135 MeV. The green-dashed curve shows the fit to the data. The other-effects emittance is shown by the blue-solid line at 0.45 microns. The fit indicates the quadrupole-solenoid emittance due to transverse coupled dynamics is 0.11 microns.

## IV. COUPLED TRANSVERSE DYNAMICS DUE TO THE ASYMMETRIC FIELDS OF RF COUPLERS

This section first discusses the asymmetric fields of an RF coupler, and that generally the fields are linear in x and y near the beam axis but rotated or skewed about the beam axis. This skewing of the beam couples the transverse dynamics which increases the emittance. Second, it is shown that the emittance generated by these coupler fields can be cancelled with the appropriately skewed quadrupole field. Since the correcting skewed quadrupole field removes the xy correlations which are characteristic of an astigmatic beam, one could refer to the corrector quadrupole as a 'quadrupole stigmator' [10].

### A. Analysis of the RF coupler field

Maxwell's equations show that the transverse electric fields over a small region near the beam axis can be specified as a linear expansion obeying the following symmetries between the field gradients,

$$E_x = E_{x,0} + \frac{\partial E_x}{\partial x} x + \frac{\partial E_x}{\partial y} y \qquad (21)$$

$$E_y = E_{y,0} + \frac{\partial E_x}{\partial y} x - \frac{\partial E_x}{\partial x} y \qquad (22)$$

A coupler gives the beam an instantaneous kick in voltage along the x, y and z-directions. And each component of the voltage kick consists of an instantaneous jump in voltage due to the transverse voltage gradient when the beam transits the coupler. Following the literature [11], the complex voltage kick factor is defined as

$$\vec{v}(x,y) = \frac{\vec{V}(x,y)}{V_z(0,0)} \cong \begin{pmatrix} v_{x0} + v_{xx}x + v_{xy}y \\ v_{y0} + v_{yx}x + v_{yy}y \\ 1 + \cdots \end{pmatrix} \qquad (23)$$

Where the complex voltage kick, $\vec{V}(x,y)$, is given by integrals of the coupler fields along lines parallel to the z-axis (beam's optical axis),

$$\vec{V}(x,y) = \int [\vec{E}(\vec{r}) + ic\vec{\beta}\times\vec{B}(\vec{r})]e^{i\omega z/c} dz \tag{24}$$

The $\vec{B}$-term is imaginary to account for its $\pi/2$ RF phase shift in time with respect to the electric field. The complex voltage kick factor gives the electrons a momentum impulse of [11]

$$\vec{p} = Re\{\vec{v}(x,y)e^{i\phi_s}\}\frac{eV_{acc}}{c} \tag{25}$$

Eqn. (25) is the transverse momentum of an electron a distance $s$ behind the head electron and having phase $\phi_s = \frac{\omega s}{c} + \phi_{head}$ with respect to the coupler's RF waveform. Writing out the components of the coupler's x-y plane momentum kick shows the spatial and phase dependences can be separated into a complex voltage amplitude and phase of which the real part is the momentum kick by the coupler,

$$\begin{pmatrix} p_x \\ p_y \end{pmatrix}_{coupler} = \frac{eV_{acc}}{c} Re\left\{\begin{pmatrix} v_{0x} + v_{xx}x + v_{xy}y \\ v_{0y} + v_{yx}x + v_{yy}y \end{pmatrix} e^{i\phi_s}\right\} \tag{26}$$

Dividing by the total momentum converts this to the coupler's angle kick,

$$\begin{pmatrix} x' \\ y' \end{pmatrix}_{coupler} = \frac{eV_{acc}}{\beta\gamma mc^2} Re\left\{\begin{pmatrix} v_{0x} + v_{xx}x + v_{xy}y \\ v_{0y} + v_{yx}x + v_{yy}y \end{pmatrix} e^{i\phi_s}\right\} \tag{27}$$

And applying the symmetry between the field gradients shown in Eqns. (21) and (22) results in coupler angle kicks with a similar symmetry between components

$$\begin{pmatrix} x' \\ y' \end{pmatrix}_{coupler} = \frac{eV_{acc}}{\beta\gamma mc^2} Re\left\{\begin{pmatrix} v_{0x} + v_{xx}x + v_{xy}y \\ v_{0y} + v_{xy}x - v_{xx}y \end{pmatrix} e^{i\phi_s}\right\} \tag{28}$$

Absorbing the various factors and phase dependence into a new voltage kick matrix given by $\tilde{v}$, allows one to write

$$\begin{pmatrix} x' \\ y' \end{pmatrix}_{coupler} = \begin{pmatrix} \tilde{v}_{0x} + \tilde{v}_{xx}x + \tilde{v}_{xy}y \\ \tilde{v}_{0y} + \tilde{v}_{xy}x - \tilde{v}_{xx}y \end{pmatrix} \tag{29}$$

The $\tilde{v}_{0x,0y}$ terms are the coupler's dipole kicks which generally can be cancelled with nearby steering dipoles. Simulations show dipole steering is very effective at mitigating the dipole-kick emittance produced when the beam goes through the coupler at the wrong angle and/or transverse position. Taking the correction concept to the next order, a quadrupole field with the proper rotation and strength can correct for the coupler's quadrupole-like kicks which generate most of the coupler emittance. Since the $\tilde{v}_{xx}$ and $\tilde{v}_{yy}$ terms are linear with respect to x and y, they affect the beam like a focusing or defocusing lens. And because this focusing is linear it does not produce any emittance. Instead the cross term, $\tilde{v}_{xy}$, or the correlation between x and y is the principal cause of emittance. The $v_{xy}y$ term of the x-kick results from the rotation of the coupler fields about the z-axis. This produces a correlation which appears as emittance in the projected xx' and yy' sub-spaces. And as a result, this correlation between the transverse coordinates looks like the rotated phase spaces just discussed for the solenoid and can be fixed with a skewed quadrupole.

### B. The emittance due to asymmetric coupler fields

For simplicity, let us assume the electron bunch distribution in the xy-plane is uniform inside a square area having dimensions, $-R < x < R$ by $-R < y < R$. This trivial distribution simplifies the calculation of the variances and averages needed for deriving the coupler-induced emittance while maintaining the essential physics. The variance and standard deviation for this *2R x 2R* square uniform distribution is,

$$\langle x^2 \rangle = \langle y^2 \rangle = \frac{R^2}{3} \quad \text{and} \quad \sigma_x = \frac{R}{\sqrt{3}} \tag{30}$$

The normalized emittance for the x-plane is defined as

$$\epsilon_n = \frac{\sqrt{\langle x^2 \rangle \langle p_x^2 \rangle - \langle xp_x \rangle^2}}{mc} = \sqrt{\langle x^2 \rangle \langle x'^2 \rangle - \langle xx' \rangle^2} \tag{31}$$

After some tedious algebra, the coupler-induced emittance is found to be solely due to the cross-term, $v_{xy}$, of the complex voltage kick,

$$\epsilon_{n,coupler}(s) = \frac{eV_{acc}}{mc^2} \sigma_x^2 \left| v_{xy}^r \cos\left(\frac{\omega s}{c} + \phi_{head}\right) + v_{xy}^i \sin\left(\frac{\omega s}{c} + \phi_{head}\right) \right| \tag{32}$$

Here $v_{xy}^r$ and $v_{xy}^i$ are the real and imaginary parts of $v_{xy}$.

Eqn. (32) gives the emittance of a thin transverse slice of the bunch, a distance *s* behind the bunch head. The RF phase of the bunch head is $\phi_{head}$ and the tail is a bunch length, $l_{bunch}$, behind it at RF phase of $\frac{\omega l_{bunch}}{c} + \phi_{head}$. Figure 9 shows as an example the head and tail emittances vs. the RF phase for a head-tail phase difference of 10 degRF. The head minus the tail emittance is also plotted and shows the difference is 20 nm or less which is small compared to the uncorrected emittance of more than 100 nm; confirming the effect is mostly due to the skewed transverse field rather than the phase-dependent kick. That is, except near 90 degRF where the bunch is on crest, the bulk of the emittance comes from the transverse spatial distribution of the coupler fields and not their time variation. Like any RF related emittance in general, the first order coupler emittance is zero on crest, leaving second-order and higher emittances due to the curvature of RF waveform.

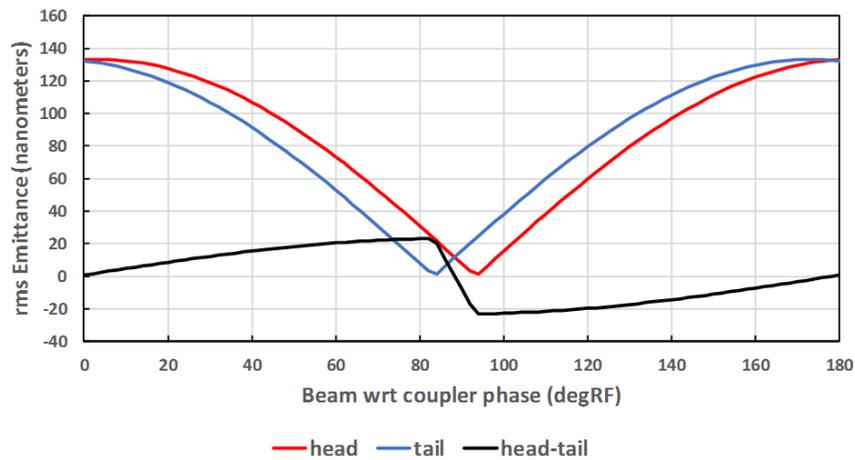

Figure 9: The head (s=0, red) and tail ($\frac{\omega s_{tail}}{c}$ =10degRF, blue) emittances vs. phase for $\sigma_x$ =1 mm. The emittance difference in the head and tail emittances is shown by the black curve. The coupler voltage and kick are V$_{acc}$=20 MV and $v_{xy} = (3.4 + 0.2i) \times 10^{-6}/mm$ which are typical parameters for SRF cavities [11].

The emittance over the length of the bunch is computed by averaging the slice emittance in Eqn. (32) over the longitudinal distribution of electrons. Assuming the longitudinal distribution is uniform with full width, $l_{bunch}$, then the bunch average emittance is given by

$$\langle \epsilon_{n,coupler} \rangle = \frac{\int_0^{l_{bunch}} \epsilon_{n,coupler}(s) ds}{\int_0^{l_{bunch}} ds} \tag{33}$$

Computing this average gives

$$\langle \epsilon_{n,coupler} \rangle = \frac{eV_{acc}}{mc^2} \sigma_x^2 \left| v_{xy}^r \left\langle \cos\left(\frac{\omega s}{c} + \phi_{head}\right) \right\rangle + v_{xy}^i \left\langle \sin\left(\frac{\omega s}{c} + \phi_{head}\right) \right\rangle \right| \tag{34}$$

And taking the averages and expanding in terms of the bunch length results in the projected emittance of the bunch,

$$\langle \epsilon_{n,coupler} \rangle = \frac{eV_{acc}}{mc^2} \sigma_x^2 \left| v_{xy}^r \cos\phi_{head} + v_{xy}^i \sin\phi_{head} \right| + O(l_{bunch}^2) \tag{35}$$

The first emittance term is independent of the bunch length, and results from the skewed, linear transverse fields of the coupler which increase the slice emittance and therefore the projected emittance as well. The $O(l_{bunch}^2)$ term gives the emittance due to the phase spread of the bunch and mostly affects the projected emittance. The first term is what can be canceled with a skewed quadrupole field.

### C. Compensation of coupler kicks with a rotated or skewed quadrupole field

A thin quadrupole lens rotated an angle $\theta_q$ about the z-axis gives the beam an angular kick very like the kick given by the coupler. This suggests using a correcting skewed quadrupole whose rotation angle and focal length can be adjusted to undo the coupler rotation, cancel its focusing strength, and thereby eliminate its emittance.

The kick angle vector in the xy-plane for a quadrupole with focal length $f_q$ and skew angle $\theta_q$ when summed with x- and y-dipole kicks can be written as

$$\begin{pmatrix} x' \\ y' \end{pmatrix}_{dipole+quad} = \begin{pmatrix} x_0' - \frac{\cos 2\theta_q}{f_q} x - \frac{\sin 2\theta_q}{f_q} y \\ y_0' - \frac{\sin 2\theta_q}{f_q} x + \frac{\cos 2\theta_q}{f_q} y \end{pmatrix} \tag{36}$$

Here it is useful to point out the similarities between the coupler kicks and those of a rotated quad. By associating terms between matrices in Eqns. (29) and (36) one finds that,

$$\tilde{v}_{xx,q} = -\frac{\cos 2\theta_{coupler}}{f_{coupler}} \quad \text{and} \quad \tilde{v}_{xy,q} = \frac{\sin 2\theta_{coupler}}{f_{coupler}} \tag{37}$$

Where the subscript 'q' has been added to the subscript to denote these are kicks of a skewed quadrupole only. These relations allow us to model the coupler fields equivalently as a quadrupole with focal length $f_{coupler}$ rotated $\theta_{coupler}$ about the z-axis. The equivalent quadrupole skew angle of the coupler field is found to be

$$\theta_{coupler} = -\frac{1}{2} \tan^{-1} \frac{\tilde{v}_{xy}}{\tilde{v}_{xx}} \tag{38}$$

While the $x_0'$ and $y_0'$ kicks are included in Eqn. (36) for completeness to explicitly show the need for dipole correctors as well as for the rotated quadrupole currently being discussed. Generally steering is available to cancel the dipole kicks, such that $x_{0,dipole}' = -\tilde{v}_{0x}$, and similarly for y. Therefore, these kick terms are ignored in what follows. And finally, the analysis will assume the longitudinal extent of the quadrupole and coupler fields is short compared to their focal lengths and located within focal lengths each other on the beamline. Then Eqns. (29) and (36) can be added (without the dipole kicks) to find the total angle kick of the coupler and the correcting quadrupole,

$$\begin{pmatrix} x' \\ y' \end{pmatrix}_{total} = \begin{pmatrix} x' \\ y' \end{pmatrix}_{coupler} + \begin{pmatrix} x' \\ y' \end{pmatrix}_{quad} = \begin{pmatrix} \left\{\tilde{v}_{xx} - \frac{\cos 2\theta_q}{f_q}\right\} x + \left\{\tilde{v}_{xy} - \frac{\sin 2\theta_q}{f_q}\right\} y \\ \left\{\tilde{v}_{xy} - \frac{\sin 2\theta_q}{f_q}\right\} x - \left\{\tilde{v}_{xx} - \frac{\cos 2\theta_q}{f_q}\right\} y \end{pmatrix} \quad (39)$$

The beauty of Eqn. (29)'s symmetry can now be appreciated since Eqn. (39) shows the emittance and the focusing effects of the coupler can be exactly eliminated by solving for the rotation and focal strength of a simple, skewed quadrupole corrector. Due to this symmetry, there are two (rather than four) equations to solve,

$$\tilde{v}_{xx} - \frac{\cos 2\tilde{\theta}_q}{\tilde{f}_q} = 0 \quad \text{and} \quad \tilde{v}_{xy} - \frac{\sin 2\tilde{\theta}_q}{\tilde{f}_q} = 0 \quad (40)$$

Simultaneously solving these two equations gives paired values for the corrector quadrupole's skew angle and focal strength which cancel the coupler field's cross-term (emittance) and the quadrupole-focus term (astigmatism) as a function of the beam-RF phase. Solving for the quadrupole skew angle and focal strength in terms of the coupler's complex voltage kick gives,

$$\tilde{\theta}_q = \frac{1}{2} \tan^{-1} \frac{\tilde{v}_{xy}}{\tilde{v}_{xx}} \quad (41)$$

$$\frac{1}{\tilde{f}_q} = \frac{eV_{acc}}{\beta \gamma mc^2} \sqrt{\tilde{v}_{xx}^2 + \tilde{v}_{xy}^2} \quad (42)$$

The normalized voltage kick factors, $\tilde{v}$, in terms of the real and imaginary parts of the usual voltage kicks, the accelerator voltage and the slice phase are,

$$\tilde{v}_{xx}(\phi_s) = \frac{eV_{acc}}{\beta \gamma mc^2} \left( v_{xx}^r \cos \phi_s - v_{xx}^i \sin \phi_s \right) \quad (43)$$

and

$$\tilde{v}_{xy}(\phi_s) = \frac{eV_{acc}}{\beta \gamma mc^2} \left( v_{xy}^r \cos \phi_s - v_{xy}^i \sin \phi_s \right) \quad (44)$$

Inserting these relations into Eqns. (43) and (44) gives the RF phase-dependent solutions for the quadrupole rotation angle,

$$\tilde{\theta}_q(\phi_s) = \frac{1}{2} \tan^{-1} \frac{v_{xy}^r \cos \phi_s - v_{xy}^i \sin \phi_s}{v_{xx}^r \cos \phi_s - v_{xx}^i \sin \phi_s} \quad (45)$$

and the quadrupole focal strength,

$$\frac{1}{\tilde{f}_{q(\phi_s)}} = \frac{eV_{acc}}{\beta \gamma mc^2} \sqrt{\left( v_{xx}^r \cos \phi_s - v_{xx}^i \sin \phi_s \right)^2 + \left( v_{xy}^r \cos \phi_s - v_{xy}^i \sin \phi_s \right)^2} \quad (46)$$

which cancel the coupler's kick. Together these two expressions satisfy Eqns. (42) and thereby eliminate the first-order coupler-induced rotation and focusing effects on the beam at the bunch-RF phase of $\phi_s$.

As a numerical example, we use the SRF coupler parameters given in Dohlus' paper [11] to compute the corrector quadrupole requirements. His Table 1 gives the normalized complex voltage kick factors as

$$v_{xx} = (1 - 0.7i) \times 10^{-6} \text{ /mm} \quad (47)$$
$$v_{xy} = (3.4 - 0.2i) \times 10^{-6} \text{ /mm} \quad (48)$$

These coupler kicks correspond to a normalizing voltage of 20 MV and assume a beam kinetic energy of 800 KeV such that $\beta\gamma = 2.5$. Figure 10 plots the focal length and skew angle required to correct for these complex voltage kicks. The correction quadrupole skew angle and focal length are plotted as functions of the bunch-head phase with respect to the coupler RF. A phase of 90 degRF corresponds to the bunch head synchronized on the RF waveform crest.

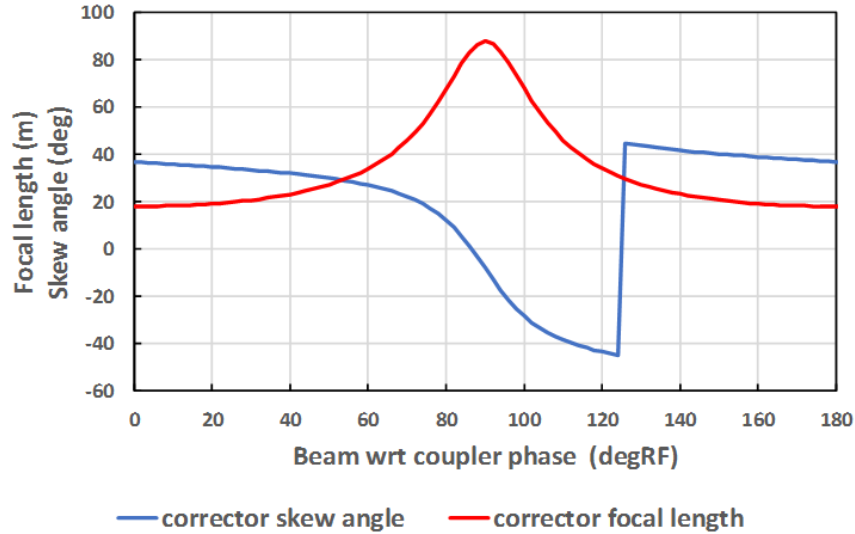

Figure 10 The correction quadrupole skew angle (blue) and focal length (red) vs. the coupler RF phase for a high-power coupler on a SRF linac. The corrector skew angle discontinuously jumps near the coupler phase of ~125 degRF where the denominator of Eq. (45) is zero and the $tan^{-1}$ function changes to stay within the its principal value range of $-\pi/2$ and $\pi/2$.

### D. Implementing quadrupole stigmators into the LCLS-II injector

LCLS-II injector consists of a CW RF gun, two solenoids for beam focusing, an RF buncher for bunch compression, and one standard 8-cavity cryomodule (CM) [8]. The standard 8-cavity CM is used to boost the beam energy to ~100 MeV from <1 MeV. The strong RF coupler's asymmetrical field located at the low energy end of the CM can increase the emittance, especially for a beam at low energy with a large beam size. Fig. 11 shows the emittance simulations for a 300-pC bunch without coupler fields (blue), with 3D coupler fields (green), and with 3D coupler fields corrected by a skewed quadrupole (red). The emittance is plotted vs z along the length of the injector to where the emittance reaches it asymptotic value. The emittance grows about 20% when the 3D RF coupler fields are included compared to the no couplers case. The simulation shows a weak quadrupole (integrated field is <2 G) located near the CM can completely correct the RF coupler's effect (red vs. green curves). Because of these studies, we are planning to use the quadrupole corrector in the second solenoid for both the stray quadrupole field and RF coupler field correction in the LCLS-II injector [12]. We may consider installing an independent quadrupole corrector near the entrance of the CM for the coupler correction if the shared quadrupole corrector is not strong enough.

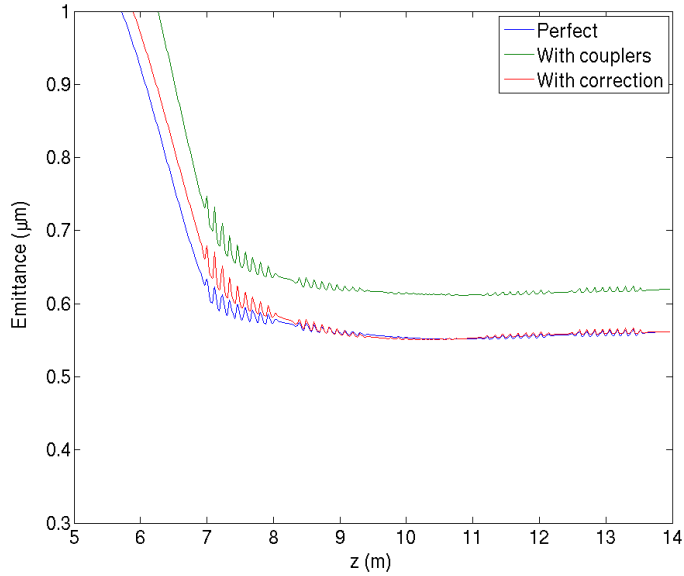

FIG. 11: LCLS-II injector emittance evolution with perfect RF field (blue), with RF couplers field (green), and correction using a weak skew quadrupole (red) for 300 pC.

## V. SUMMARY AND CONCLUSIONS

This paper discussed two types of emittance due to coupled transverse dynamics. These emittances are caused by skewed fringe quadrupole fields of the solenoid and their interaction with the solenoid field, and the skewed RF quadrupole fields of high power couplers for accelerator linacs. Both effects are commonly encountered in high brightness guns and injectors.

The emittance increase caused by a weak quadruple field in combination with a solenoidal field is analyzed for electron injectors. And its correction with a weak skewed quadrupole is verified through analytical theory, simulations, and experiments. It is also shown that RF couplers in RF cavities can significantly degrade the emittance. Further theoretical analysis and simulations show that the emittance growth can be completely corrected with a weak skewed quadrupole. These quadrupole correction techniques for solenoids and RF couplers have been implemented into the LCLS-I and LCLS-II injectors as well as in the SwissFEL injector [13] and the Cornell DC photocathode injector [14].

In addition to the stray quadrupole fields, we have begun studies of the emittance due to stray sextupole fields. Preliminary results indicate this is also a potential source of emittance, especially for large size beams in a sextupole field. This is because the sextupole emittance scales with the beam size to the third-power. Such large beams are typical in guns and injectors which operate at high-duty factor but low peak field. Therefore, continued studies of skewed sextupole fields in conjunction with solenoid focusing are needed to explore correcting with weak sextupole correctors.

And finally, there is the potential use of skew quadrupole correctors in highly dispersive systems such as chicane bunch compressors to correct for small coupling of the transverse dynamics between the energy dispersed plane and the orthogonal, non-dispersed plane dynamics due to slightly rotated dipoles. These and other examples demonstrate the general applicability of skewed quadrupoles to uncouple the transverse dynamics of electron beams.

**ACKNOWLEDGEMENTS**

The work is supported by DOE under grant No. DE-AC02-76SF00515.